\documentclass[fleqn,usenatbib]{mnras}

\usepackage{newtxtext,newtxmath}

\usepackage[T1]{fontenc}
\usepackage{ae,aecompl}

\usepackage{graphicx}	
\usepackage{amsmath}	
\usepackage{amssymb}	
\usepackage{xcolor}
\usepackage{hyperref}

\title[Effects of the magnetic field direction on the Tsallis statistic]{Effects of the magnetic field direction on the Tsallis statistic}

\author[Gonz\'alez-Casanova, Lazarian \& Cho]{
Diego F. Gonz\'alez-Casanova,$^{1}$\thanks{E-mail: casanova@astro.wisc.edu}
A. Lazarian,$^{1}$\thanks{E-mail: lazarian@astro.wisc.edu}
J. Cho$^{2}$
\\
$^{1}$Astronomy Department, University of Wisconsin-Madison, 475 North Charter Street, Madison, WI 53706-1582, USA\\
$^{2}$Department of Astronomy and Space Science, Chungnam National University, Daejeon, Korea\\
}

\date{Accepted XXX. Received YYY; in original form ZZZ}

\pubyear{2015}

\begin{document}
\label{firstpage}
\pagerange{\pageref{firstpage}--\pageref{lastpage}}
\maketitle

\begin{abstract}
We extend the use of the Tsallis statistic to measure the differences in gas dynamics relative to the mean magnetic field present from natural eddy-type motions existing in magnetohydrodynamical (MHD) turbulence. The variation in gas dynamics was estimated using the Tsallis parameters on the incremental probability distribution function of the observables (intensity and velocity centroid) obtained from compressible MHD simulations. We find that the Tsallis statistic is susceptible to the anisotropy produce by the magnetic field, even when anisotropy is percent the Tsallis statistic can be use to determine MHD parameters such as the Sonic Mach number. We    quantize the goodness of the Tsallis parameters using the coefficient of determination to measure the differences in the gas dynamics. These parameters also determine the level of magnetization and compressibility of the medium. To further simulate realistic spectroscopic observational data we introduced smoothing, noise, and cloud boundaries to the MHD simulations.  

\end{abstract}



\section{Introduction}

The dynamics of the interstellar medium (ISM) regulate various phenomena such as star formation, cosmic ray physics, galaxy evolution, and magnetic dynamo theory.  Two fundamental constituents of the ISM are magnetic fields and turbulence \citep[see reviews by][]{elmegreen2004, maclow2004, mckee2007}, making it essential to study the properties of magnetohydrodynamical (MHD) turbulence. Essential evidence of MHD turbulence in the ISM is seen in the energy power law \citep{chepurnov2010, armstrong1995}, fractal structure in the molecular media \citep{elmegreen1996, stutzki1998}, and in intensity fluctuations of both density and turbulent velocity \citep{crovisier1983, green1993, elmegreen2001}. 

The most used tool to analyze the turbulent properties of the ISM is the power spectrum calculation for both observational and computational approaches (where the slope of the power-law determines some of the properties of the medium).  These calculations have been done on scintillation fluctuations \citep{narayan1989}, density fluctuations via column density \citep{falgarone1994, lis1998, miesch1999} and radio spectroscopy \citep{lithwick2001, cho2003}. The power index can also be estimated from position-position-velocity (PPV) data via Velocity Channel Analysis \citep[VCA][]{lazarian2000, padoan2003, kandel2016}, Spectral Correlation Function \citep{rosolowsky1999, padoan2001}, and Velocity Coordinate Spectrum \citep[VCS][]{lazarian2006b, lazarian2008b, padoan2009}, among others.

The power spectrum is only one of the properties that the turbulent medium has. Other parameters include sonic and Alfv{\'e}n Mach numbers ($M_s$ and $M_A$), injection scale, gas temperature, and Reynolds number. In particular, the sonic and Alfv{\'e}n Mach numbers provide information on the gas compressibility and magnetization, two properties that are convoluted in the power index.  
 
The properties of the ISM are such that the \citet{tsallis1988} statistics have been used to analyze its compressibility and magnetization, with \citet{esquivel2010} being the first ones to apply it to MHD simulations.  The statistic was first used in an astronomical context by \citet{burlaga2004} to prove the properties of the solar wind, using information obtained from the Voyager missions. \citet{esquivel2010} fit the Tsallis distribution to incremental probability distribution functions (PDF) of the 3D properties of the medium (density, velocity and magnetic field), finding good correlation between the Tsallis distribution and its $M_s$ and $M_A$. \citet{tofflemire2011} applied a similar procedure to observable quantities such as the intensity and the projected magnetic field.  This analysis was able to obtain reliable estimates of $M_s$ and $M_A$, but with quantities obtained from spectroscopic observations.  Therefore the Tsallis statistic is a reliable technique to understand the turbulent properties of the ISM. 

Turbulent MHD has a key difference from purely hydrodynamical turbulence: the eddy motions align with the local magnetic field \citep{goldreich1995,lazarian1999}. The motions perpendicular and parallel to the mean magnetic field are therefore different. The differences in motion are seen in density and velocity structures \citep[e.g.][]{esquivel2005, soler2013,kandel2016}. In this paper we extend the use of the Tsallis statistic to MHD simulations to analyze and quantify the effects of the magnetic field in the fluid motions relative the the mean magnetic field. We apply it to observables (intensity and velocity centroids) and add different effects to simulate realistic observations. In Section~\ref{sec:theo}, we explore the characteristics of the Tsallis statistic; in Section~\ref{sec:num}, we describe the numerical code and setup for the MHD simulations; in Section~\ref{sec:results}, we present our results on the anisotropy by means of the Tsallis distribution; in Section~\ref{sec:discussion}, we discuss our results; and in Section~\ref{sec:conclusion}, we give our conclusions.


\section{Theoretical Considerations - The Tsallis Statistic}\label{sec:theo}

Complex systems such as turbulent gas dynamics require a formalism known as non-extensive statistical mechanics.  These systems have the characteristic of non-Gaussian distributions with asymptotic power law behavior, long range correlations and multifractality \citep{picoli2009}.  \citet{tsallis1988} developed the Tsallis statistic as a generalization of the entropy in fractal systems.  The Tsallis distribution, or q-Gaussian distribution ($P_q(\Delta f(r))$), is a generalization of the Gaussian distribution that complex systems satisfy:  

\begin{equation}
P_q(\Delta f(r)) = a \Big[ 1+(q-1)\frac{\Delta f(r)^2}{w^2} \Big]^{-1/q-1} \;,
\label{eq:tsallis}
\end{equation}
where $a$, $q$ and $w$ are the Tsallis parameters and $\Delta f(r)$ corresponds to the incremental PDF in the space of the fitted function ($f$).  The different parameters modify the structure of the distribution: $q$ controls the tail, $w$ the width, and $a$ the amplitude of the distribution. The incremental PDF of $f$ fitted by the Tsallis distribution is defined as:
 
\begin{equation}
\Delta f(r) = \frac{f (r)-\langle f (r) \rangle_x}{\sigma_f} \;, 
\label{eq:fit}
\end{equation}
where $\langle$...$\rangle_x$ refers to a spatial average; $\sigma_f$ to its respective standard deviation; and $r$ the `lag' referring to the spatial variation of the data points across the data set. Both $\langle$...$\rangle_x$ and $\sigma_f$ are a function of the lag.


For this work we used observables ($I$, $S$ and $C$) as the input for the Tsallis statistic.  The observables are the projected quantities into the plane of the sky along the line of sight (LOS). Velocity centroids $ C(\mathbf{x})$ and $S(\mathbf{x})$ give information on the velocity field of the medium.  The intensity $I(\mathbf{x})$ depends on the emitting source and can be proportional to the density, such as in the case of HI emission.  Calculation of the velocity centroids require spectroscopic information.  The observables are defined as:

\begin{eqnarray}
C(\mathbf{x}) \propto \frac{ \int v_z(\mathbf{x},z) \rho(\mathbf{x},z) dz }{\int \rho(\mathbf{x},z) dz}\;, \hfill \nonumber \\
S(\mathbf{x}) \propto \int v_z(\mathbf{x},z) \rho(\mathbf{x},z) dz\;, \hfill \nonumber \\
I(\mathbf{x}) \propto \int \rho(\mathbf{x},z) dz\;, \hfill \nonumber \\
\label{eq:centroid}
\end{eqnarray}
where $\rho$ is the density, $v_z$ is the LOS component of the velocity, $\mathbf{x}$ is the position of the plane of the sky, $z$ the position along the LOS,  $C(\mathbf{x}) $ is the normalized velocity centroid, and $S(\mathbf{x}) $ the un-normalized velocity centroid. The velocity centroids and the intensity for this work do not take into account self-absorption and assume an optically thin medium.  Therefore the incremental PDF for the observables, $f(r)$, that the Tsallis distribution fits takes the from of: $f(r) = C(\mathbf{x+r})-C(\mathbf{x})$; $f(r) = S(\mathbf{x+r})-S(\mathbf{x})$; and $f(r) = I(\mathbf{x+r})-I(\mathbf{x})$, where $\mathbf{r}$ is the lag and $\mathbf{x}$ is the position in the plane of the sky.  Because the lag is in vector form we can construct different functions that take the lag to be  parallel, perpendicular, and isotropic (both directions) relative to the mean magnetic field.

\section{MHD Simulations}\label{sec:num}

We use an isothermal ideal compressible MHD code \citep[see][for details]{cho2002b}. The code solves the ideal MHD equations with periodic boundary conditions with a third-order hybrid, essentially non-oscillatory (ENO) scheme. The turbulence is driven solenoidally in Fourier space with a wave scale 2.5 times smaller than the simulation box size, i.e. $2.5~l_{inj} = l_{box}$. The simulation is run until the energy reaches saturation from the turbulence driving.  The code solves the problem in code units (scale-free) so they can be scaled for any parameters of the different media in the ISM \citep{draine1998, burkhart2009}. The initial density and velocity are set to unity, length is defined in terms of the box size ($L$), and the time as the eddy turnover time ($L/ \delta v$). To reach saturation and stability, the simulations runs for 5-7 eddy turnover times until the RMS velocity is close to $v_{rms}$~$\sim$~0.7.  The sonic Mach number is defined as: $M_S = \langle |v_{rms}|/c_s \rangle$, and the Alfv{\'e}n Mach number is defined as: $M_A = \langle |v_{rms}|/V_A \rangle$, where $c_s$ is the sound speed and $V_A$ is the Alfv{\'e}n speed.  

The magnetic field has a uniform initial contribution ($\mathbf{B_o}$) in the `$x$' direction perpendicular to the LOS. Once the simulation reaches convergence, the magnetic field has a mean component and a fluctuating one i.e. ($\mathbf{B} = \mathbf{B_o}+\mathbf{\delta b}$).  Two cases are simulated: $\mathbf{B_o} =1$ (sub-Alfv{\'e}nc; $M_A\approx0.7$) and $\mathbf{B_o} =0.1$ (super-Alfv{\'e}nic; $M_A\approx2$). Initially $\mathbf{\delta b}=0$ and for all times $\langle \mathbf{\delta b} \rangle \approx 0$.  The fluctuating field is produced by the turbulent motion and energy injection.  Our data set consists of 18 numerical simulations with a resolution of $512^3$ with different Mach numbers (see Table \ref{table:data}). 

\begin{table}
\centering
\caption{Simulation parameters}
\label{table:data}
\begin{tabular}{lrrrrl}
\hline
\multicolumn{1}{c}{Model} & P$_{\rm{ext}}$& $B_{o}$ & $M_s$ & $M_A$ & Description                   \\ \hline \hline
1   & 0.0049       & 0.1     & 10    & 2.0   & Supersonic and super-Alfv{\'e}nic \\ 
2   & 0.0049       & 1.0     & 10    & 0.7   & Supersonic and sub-Alfv{\'e}nic   \\
3   &0.0077        & 0.1     & 7     & 2.0   & Supersonic and super-Alfv{\'e}nic \\
4   &0.0077        & 1.0     & 7     & 0.7   & Supersonic and sub-Alfv{\'e}nic   \\
5   &0.01            & 0.1     & 6     & 2.0   & Supersonic and super-Alfv{\'e}nic \\
6   &0.01            & 1.0     & 6     & 0.7   & Supersonic and sub-Alfv{\'e}nic   \\ 
7   &0.025          & 0.1     & 5     & 2.0   & Supersonic and super-Alfv{\'e}nic \\
8   &0.025          & 1.0     & 5     & 0.7   & Supersonic and sub-Alfv{\'e}nic   \\
9   &0.05            & 0.1     & 4     & 2.0   & Supersonic and super-Alfv{\'e}nic \\
10   &0.05          & 1.0     & 4     & 0.7   & Supersonic and sub-Alfv{\'e}nic   \\
11   &0.1            & 0.1     & 3     & 2.0   & Supersonic and super-Alfv{\'e}nic \\
12 &0.1              & 1.0     & 3     & 0.7   & Supersonic and sub-Alfv{\'e}nic   \\
13 &0.7              & 0.1     & 1     & 2.0   & Subsonic and super-Alfv{\'e}nic   \\
14 &0.7              & 1.0     & 1   	& 0.7   & Subsonic and sub-Alfv{\'e}nic     \\
15 &1                 & 0.1     & 0.7   & 2.0   & Subsonic and super-Alfv{\'e}nic   \\
16 &1                 & 1.0     & 0.7   & 0.7   & Subsonic and sub-Alfv{\'e}nic     \\
17 &2                 & 0.1     & 0.1   & 2.0   & Subsonic and super-Alfv{\'e}nic   \\
18 &2                 & 1.0     & 0.1   & 0.7   & Subsonic and sub-Alfv{\'e}nic    \\ \hline
\end{tabular}
\end{table}


\section{Tsallis fit of the intensity and the velocity centroids} \label{sec:results}

\begin{figure}
\centering\includegraphics[width=\linewidth,clip=true]{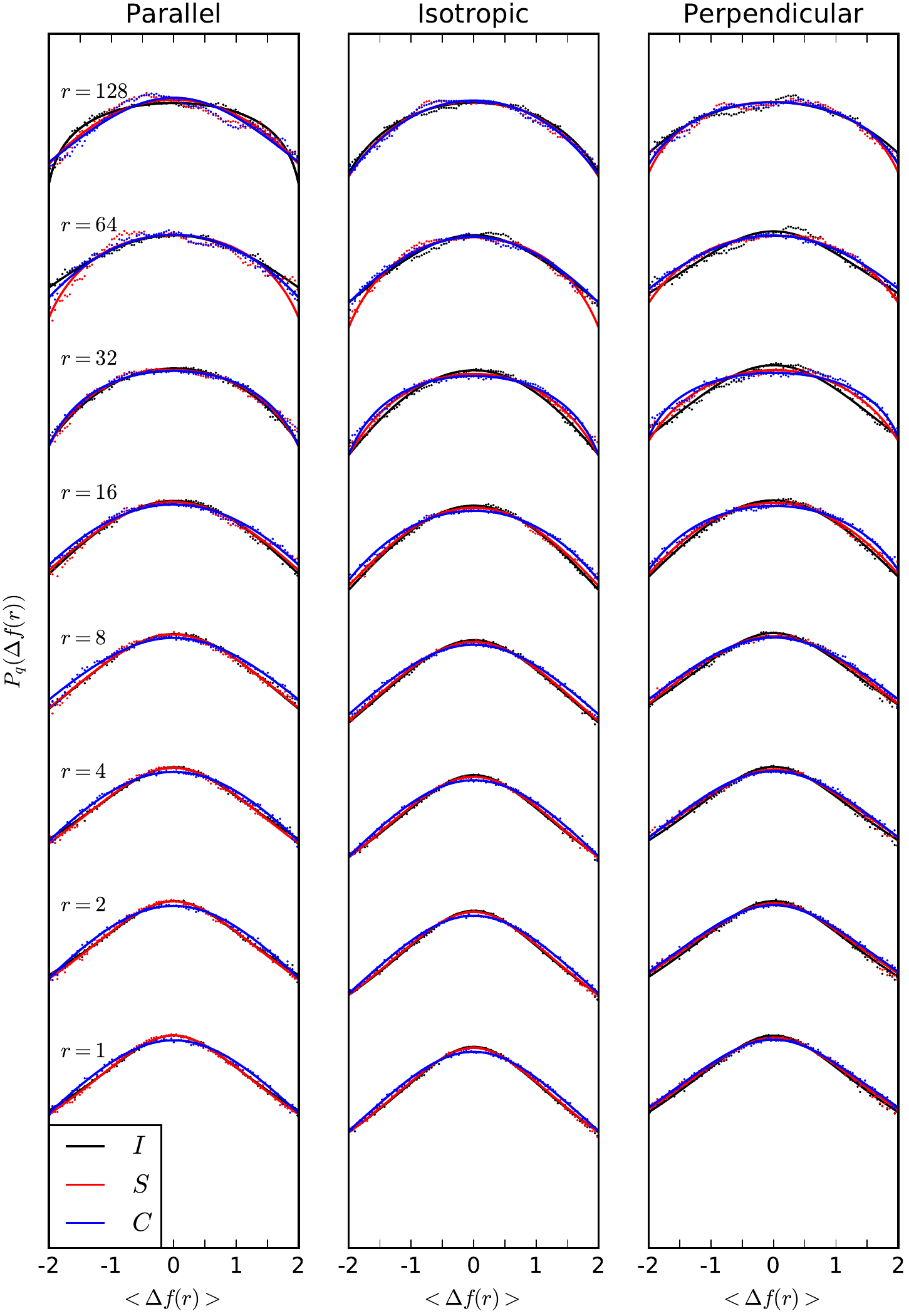}
\caption{The incremental PDF (data points and fitted Tsallis function) for the three observables (un-normalized centroid {\it red}, normalized centroid {\it blue} and intensity {\it black}) for Model 14 ($M_s$=1.0 and $M_A$=0.7; supersonic and super-Alfv{\'e}nic). The PDF shows eight different lags (1, 2, 4, 8, 16, 32, 64, 128) displayed along the `$y$' axis. The `$y$' axis is log scale.  The fitting process is done parallel ({\it left panel}), perpendicular ({\it right panel}), and isotropically ({\it center panel}) with respect to the mean magnetic field.}
\label{fig:fit}
\end{figure}

We fit the Tsallis distribution (equation \ref{eq:tsallis}) to the incremental PDFs of the two velocity centroids ($C$ and $S$) and the intensity ($I$) using the Levenberg-Marquardt algorithm \citep{levenberg1994, marquardt1963}.  The fitting process is done to individual components (parallel and perpendicular) and to both components (isotropic). Figure \ref{fig:fit} shows the data points and the fitted function for model 14 (supersonic and super-Alfv{\'e}nic). The fitting process for all cases is done to the incremental PDF using 100 bins and using the range from -2 to 2.  The lag was changed from 1 to 128 in powers of 2 for all fits.

\begin{figure*}
\centering\includegraphics[width=\linewidth,clip=true]{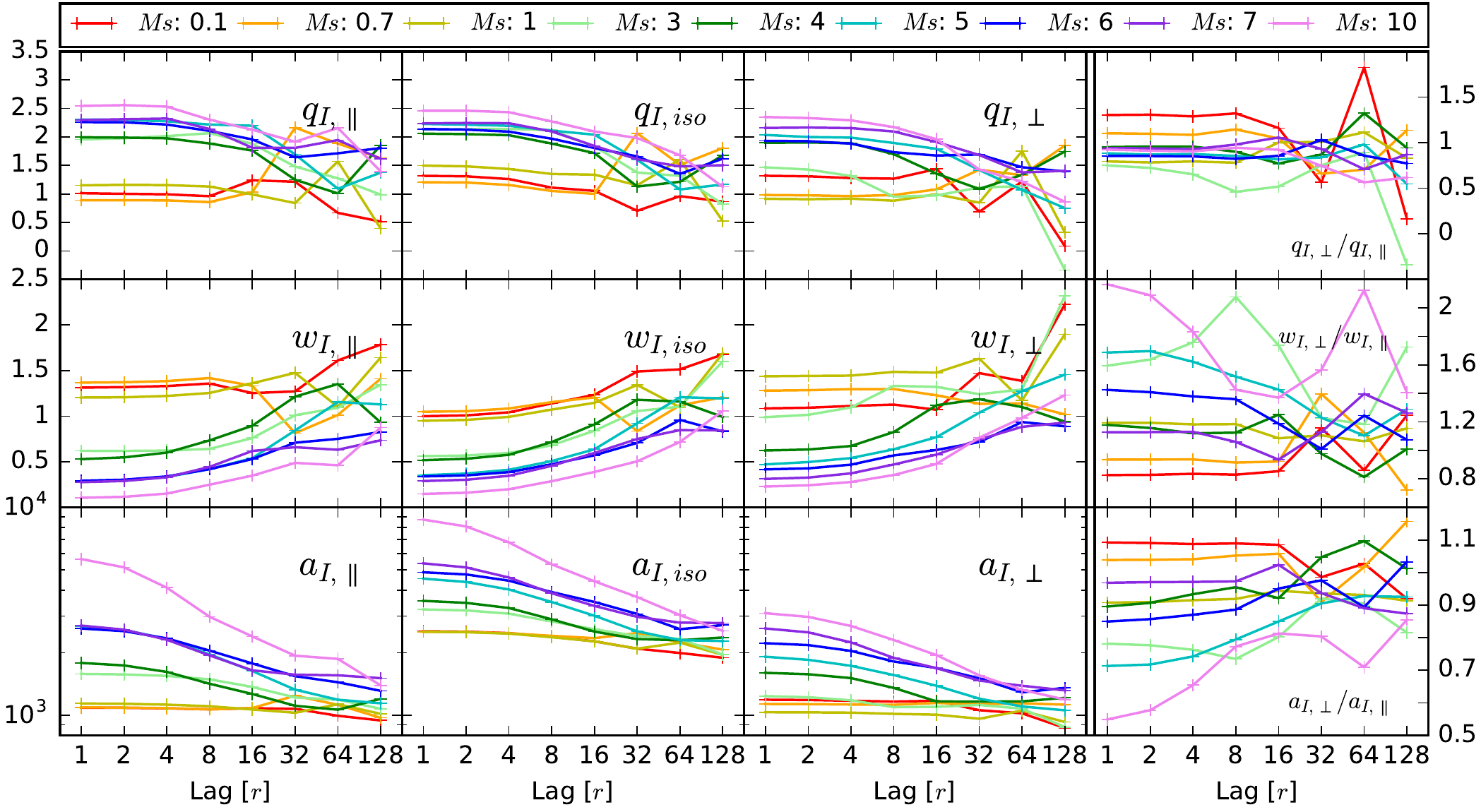}
\caption{The Tsallis parameters, $q_i$ ({\it Top row}), $w_i$ ({\it middle row}), and $a_i$ ({\it bottom row}) for the sub-Alfv{\'e}nic regime for the intesity. The parameters are shown as a function of the lag ($r$), for the nine different $M_s$ (color coded).  The first left three columns correspond to the parallel, isotropic and perpendicular cases.  The right column corresponds to the ratio of the parallel and perpendicular Tsallis parameters, also as a function of the lag.}  
\label{fig:param-de-b1}
\end{figure*}

\begin{figure*}
\centering\includegraphics[width=\linewidth,clip=true]{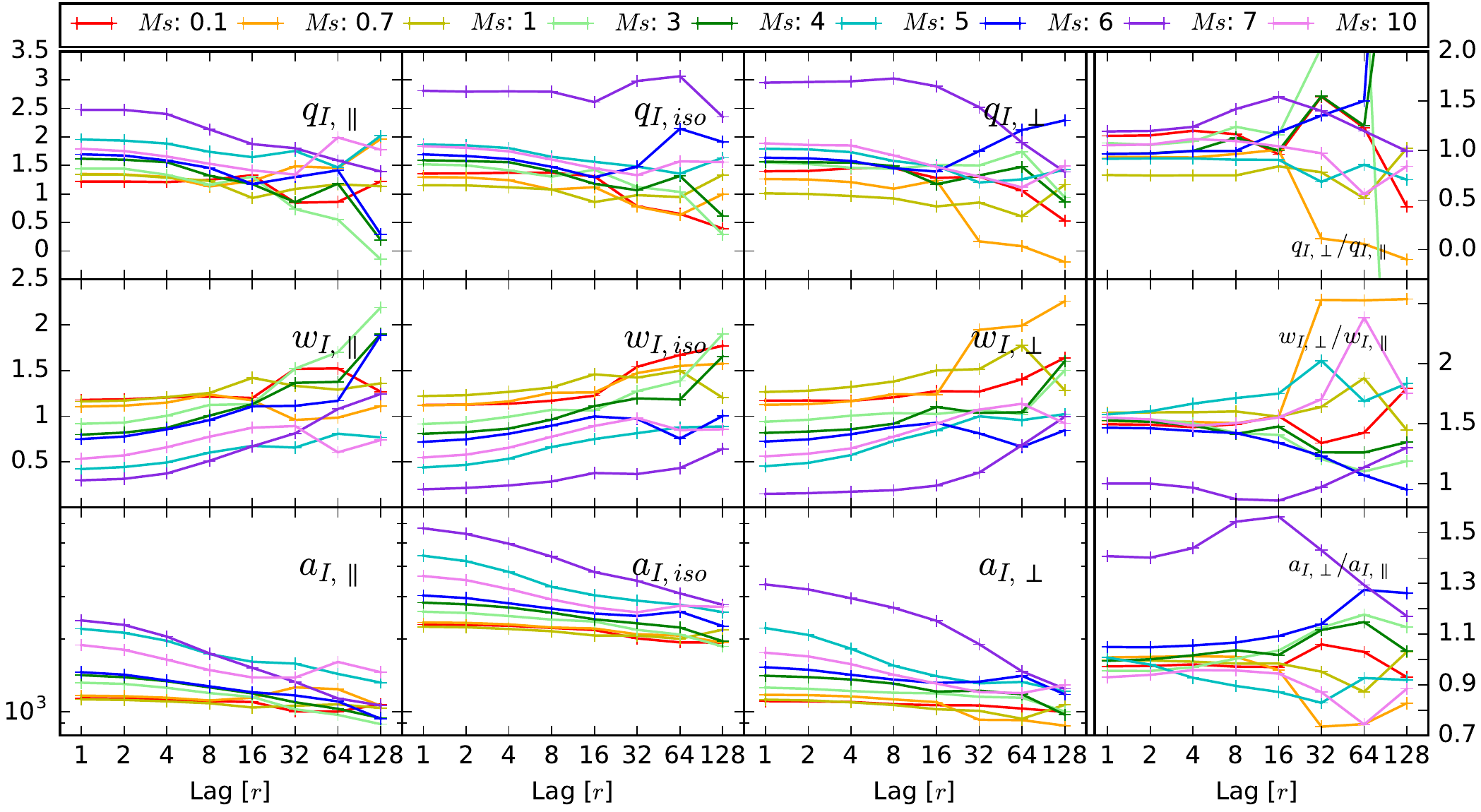}
\caption{The Tsallis parameters, $q_i$ ({\it Top row}), $w_i$ ({\it middle row}), and $a_i$ ({\it bottom row}) for the super-Alfv{\'e}nic regime for the intesity. The parameters are shown as a function of the lag ($r$), for the nine different $M_s$ (color coded).  The first left three columns correspond to the parallel, isotropic and perpendicular cases.  The right column corresponds to the ratio of the parallel and perpendicular Tsallis parameters, also as a function of the lag.}
\label{fig:param-de-b01}
\end{figure*}

\begin{figure*}
\centering\includegraphics[width=\linewidth,clip=true]{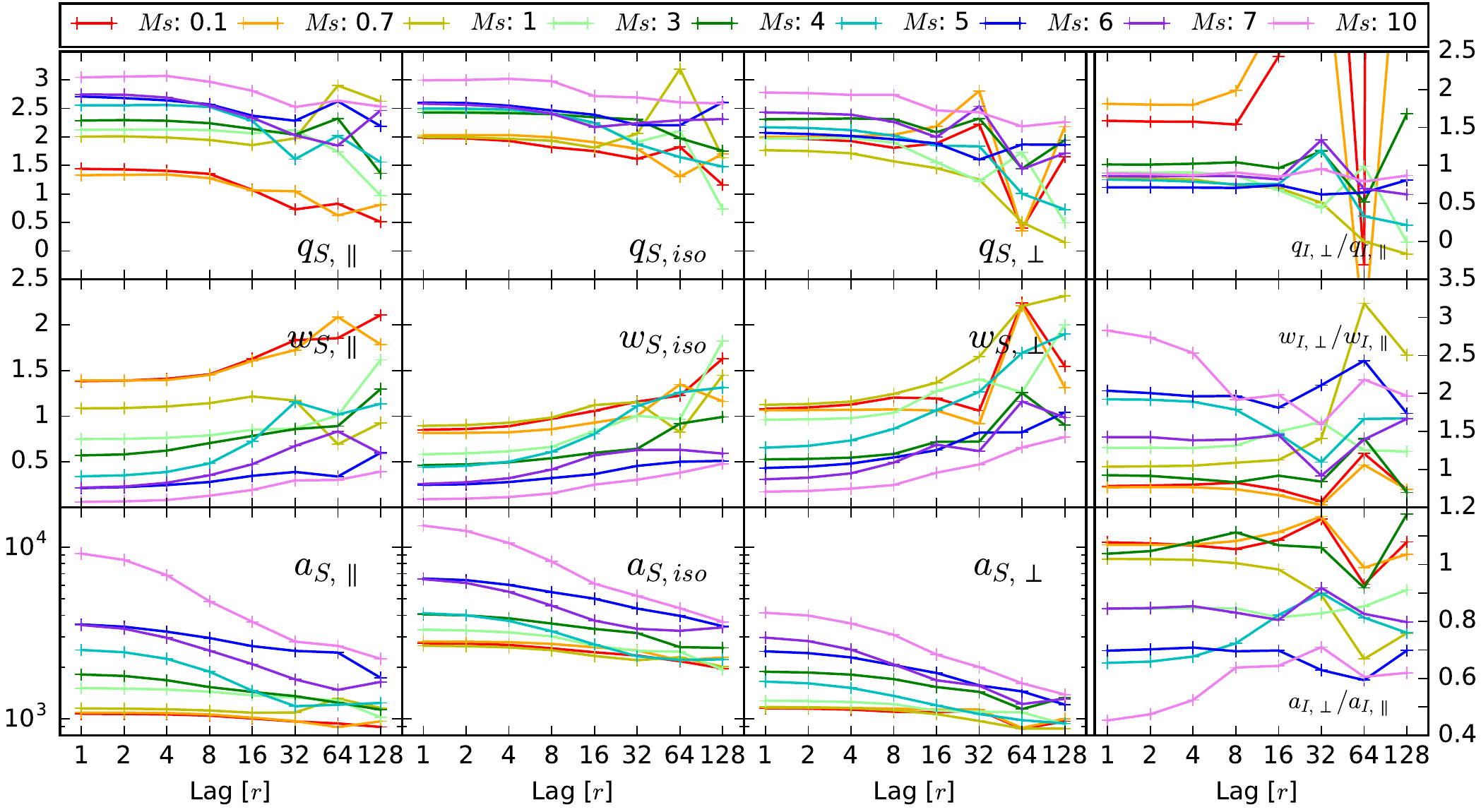}
\caption{The Tsallis parameters, $q_i$ ({\it Top row}), $w_i$ ({\it middle row}), and $a_i$ ({\it bottom row}) for the sub-Alfv{\'e}nic regime for the un-normalized velocity centroid. The parameters are shown as a function of the lag ($r$), for the nine different $M_s$ (color coded).  The first left three columns correspond to the parallel, isotropic and perpendicular cases.  The right column corresponds to the ratio of the parallel and perpendicular Tsallis parameters, also as a function of the lag.} 
\label{fig:param-uc-b1}
\end{figure*}

\begin{figure*}
\centering\includegraphics[width=\linewidth,clip=true]{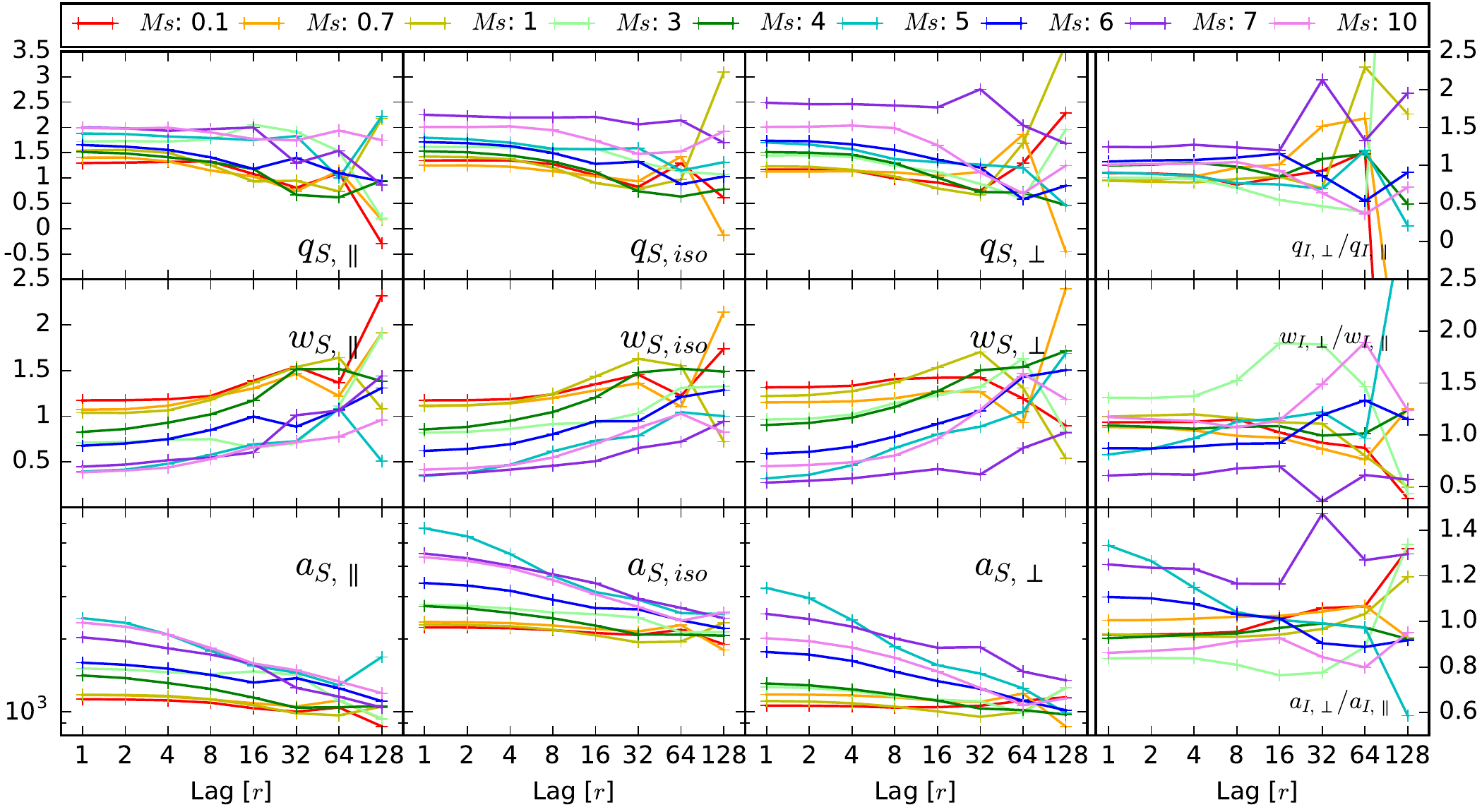}
\caption{The Tsallis parameters, $q_i$ ({\it Top row}), $w_i$ ({\it middle row}), and $a_i$ ({\it bottom row}) for the super-Alfv{\'e}nic regime for the un-normalized velocity centroid. The parameters are shown as a function of the lag ($r$), for the nine different $M_s$ (color coded).  The first left three columns correspond to the parallel, isotropic and perpendicular cases.  The right column corresponds to the ratio of the parallel and perpendicular Tsallis parameters, also as a function of the lag.} 
\label{fig:param-uc-b01}
\end{figure*}

For each of the models (Table \ref{table:data}) the Tsallis fit gives nine parameters for each of the three observables and three parameters for each of the fits ($a_{iso}$, $a_\perp$, $a_\|$, $q_{iso}$, $q_\perp$, $q_\|$, $w_{iso}$, $w_\perp$, $w_\|$) that are lag-dependent. In total there are 27 parameters for each of the 18 models.  Figures \ref{fig:param-de-b1} and \ref{fig:param-uc-b1} present the $a$, $q$, and $w$ parameters for the sub-Alfv{\'e}nic regime for the intensity and the un-nomalized centroid respectively.  Figures \ref{fig:param-de-b01} and \ref{fig:param-uc-b01} present the same parameters for the super-Alfv{\'e}nic regime for the intensity and the un-nomalized centroid respectively.  The Tsallis parameters for the normalized velocity centroid are not shown, but have a similar behavior to the un-nomalized one. The panel that shows the isotropic case in Figures \ref{fig:param-de-b1}-\ref{fig:param-uc-b01} reproduces the results found in \citet{tofflemire2011}.

The ability to distinguish the differences in the motions, if such, has to be reflected in the Tsallis parameters. Figures \ref{fig:param-de-b1}-\ref{fig:param-uc-b01} on the last column shows the ratio of the three Tsallis parameters for the two directions. Since the ratio is different than one, the effect of the mean magnetic field in the media is measurable. As seen, the sub-Alfv{\'e}nic regimen presents a stronger effect in the medium than the super-Alfv{\'e}nic case as one would expect, for both the intensity and the velocity centroid.  The Tsallis parameters can then be used to determine both Mach numbers using the different figures. 

The Tsallis parameters measure the differences in the dynamics, but the parameters are only as good as the Tsallis distribution fits the properties of the medium.
It is therefore important to estimate the goodness of fit of the Tsallis function on the incremental PDF.  This estimate is related to the errors of the Tsallis parameters, and thus to their ability to determine the effects of the mean magnetic field.  To estimate the goodness of fit we use the coefficient of determination, $R^2$:

\begin{equation}
R^2 = 1- \frac{\sum_i^N (f_i-\bar{y})^2}{\sum_i^N(y_i-\bar{y})^2} \;,
\label{eq:r_squared}
\end{equation}
where $N$ is the number of points of the data set (in this case the number of bins), $f_i$ is the model data from the Tsallis fit, and $y_i$ is the data points from the MHD simulations.  A value of one in the coefficient of determination means the model fits the data points perfectly. Therefore the closer $R^2$ is to one, the better the model fits the data points.

\begin{figure*}[h]
\centering\includegraphics[width=\linewidth,clip=true]{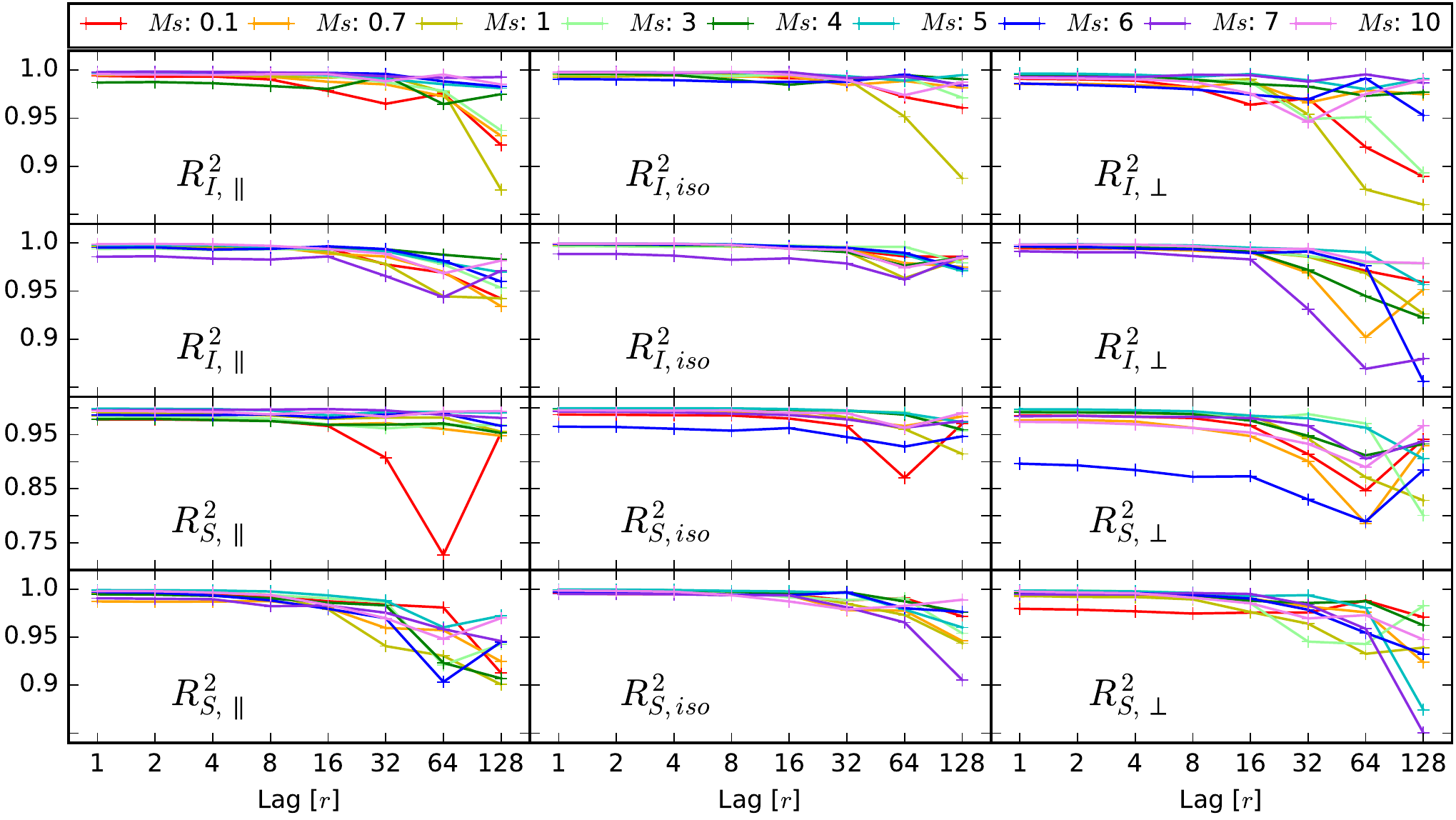}
\caption{ The coefficient of determination ($R^2$) for the intensity ({\it top two rows}) and for the un-normalized centroid ({\it bottom two rows}) for the sub-Alfv{\'e}nic ({\it first and third rows}) and super-Alfv{\'e}nic regimes  ({\it second and forth}) as a function of the lag and the sonic Mach number.  {\it Left column} for the parallel, {\it center column} for the isotropic and {\it right column} for the perpendicular cases.}
\label{fig:r_square}
\end{figure*}

The coefficient of determination was estimated for all 18 MHD models with 8 lags in the 3 directions (see Figure \ref{fig:r_square}). Table \ref{table:R} shows the value of the $5^{th}$ percentile, i.e. the  value of $R^2$ above which 95\% of the data lay.  We chose the $5^{th}$ since is shows the accuracy of the overall technique and disregards outliers. The global minimum is also reported, clearly showing that for all cases the coefficient is close to one, implying an adequate fit between the Tsallis distribution and the incremental PDF of the observables. 

Figure \ref{fig:r_square} also shows that the larger the lag the lower the values of the coefficient, or in other words the poorer the correlation becomes. We estimated $R^2$ but now for data with a lag smaller than 32 (see Table \ref{table:R}).  The lag of 32 is chosen a lag, that increases $R^2$ without limiting the range of observations. This approach significantly increases the value of the correlation parameter, translating to a more accurate measure of the effects of the magnetic field.

\begin{table}
\centering
\label{table:R}
\begin{tabular}{c|cccc|ccc}
       & \multicolumn{4}{|c|}{Full lag range} & \multicolumn{3}{c|}{1-32 lag range} \\
       & $\perp$  & $iso$  & $\|$   & min    & $\perp$     & $iso$     & $\|$      \\ \hline
$I(x)$ & 0.905    & 0.945  & 0.944  & 0.856  & 0.982       & 0.987     & 0.984     \\
$S(x)$ & 0.871    & 0.905  & 0.923  & 0.727  & 0.919       & 0.971     & 0.975    
\end{tabular}
\caption{The coefficient of determination for the two observables.  The first four measurements correspond to the full lag range and the last three columns to a lag between 1-32.  The values reported for the perpendicular, parallel and isotropic measurements correspond to the value at the $5^{th}$ percentile.  The min value column is the minimum value of $R^2$ that it was found for its respective observable in the full lag range.}
\end{table}

\section{Discussion} \label{sec:discussion} 

Understanding turbulent gas dynamics in the ISM, particularly the magnetization and compressibility, is fundamental to understanding star formation, the phases of the ISM, and cosmic ray transport, among others \citep{mckee1977,audit2005,balsi2013,li2014}. Without magnetic fields in the presence of turbulence, several astronomical phenomena could not be explained, such as star formation \citep{hennebelle2009a,santos2010}. Because cloud conditions such as the sonic and Alfv{\'e}nic Mach numbers ($M_S$ and $M_A$) affect the initial properties and outcomes of star-forming regions, it is vital to better characterize the medium \citep{burkhart2013}.  The magnetic field direction, which directly affects gas motions and star formation, is additionally vital to comprehend.

We use the Tsallis statistic applied to MHD simulations to directly determine the differences in gas motions imposed by the magnetic field. This is seen in the ratio of the different parameters shown in the last column of Figures \ref{fig:param-de-b1}-\ref{fig:param-uc-b01}.  Using the coefficient of determination we infer that the Tsallis parameters can properly describe the anisotropy of the medium.  The degree of anisotropy is better measured at smaller lag.  This is in accordance with current MHD theory, where the eddy motions that produce the anisotropy have specific scaling laws that imply that the smaller the size of the eddy, the higher the anisotropy of the eddy \citep{goldreich1995,lazarian1999,cho2002}.

The $q$ and $w$ parameters control the shape of the distribution.  Based on the particular properties that they control, these parameters can be related to the standard deviation and kurtosis (the second- and fourth- order statistical moment of a normal distribution).  The main difference is that $q$ and $w$ are not defined independently. In the same way that kurtosis measures the effects of Mach numbers, so can the $q$ parameter, as seen in the respective figures \citep{kowal2007,burkhart2009}.

Realistic observational data has intrinsic inefficiencies that decrease the quality of the data and therefore could lower the effectiveness of the Tsallis statistic to understand the motions present in the ISM. We incorporate the effects of smoothing, noise and cloud boundaries effects in the next subsections.

\subsection{Data smoothing}

Observational spectroscopic data does not have a pencil-thin beam like computer simulations but rather a smooth beam resolution.  In order to simulate the effects of telescope data we use a gaussian smoothing kernel, where we change the full width at half maximum (FWHM).  The FWHM is measured in number of cells. We use four values for the FWHM (2, 4) to smooth the data for the three observables.  Larger values for the FWHM producs $R^2$ numbers that are too small (less than 0.85). 

The smoothing procedure changes the values of the Tsallis parameters compared to the values with no smoothing, but the overall trend is preserved. Even with smoothing, the ratio of the Tsallis parameters reflect the anisotropy of the media. Similarly to the un-smoothed data, the parameters should not be measured at lags larger than 32.  The coefficient of determination is similar between the two datasets (smoothed and un-smoothed), confirming a good measurement even with smooth data.

\subsection{Noise}

Realistic observational data has intrinsic noise. To simulate this, we added artificial noise to the simulations.  We introduced Poisson noise to the projected data (the plane of the sky) with signal to noise ratio (S/N) values of 10, 50, 100, and 400.  The noise is added independently to both the intensity and velocity maps.  The noise changes the values of $I(x)$ and $S(x)$, which then changes the value of the Tsallis parameters. Using the same procedure described in section \ref{sec:results} we obtained the Tsallis parameters.

The coefficient of determination was estimated for both Alfv{\'e}nic regimes, and in both cases the $5^{th}$ percentile of the fitted models have an $R^2$ above 0.933 for the intensity and above 0.892 for the un-normalized centroid.  The Tsallis parameters, like in the case of smoothing, vary from the pure signal data, but even then the variations are small and they properly fit the data.  Therefore one is able to measure differences in gas dynamics even with noisy data. 

\subsection{Cloud effects}

All previous analysis corresponds to diffuse media observations of the ISM, but ISM observations can also include clumpy objects such as molecular clouds. To simulate a cloud we apply a gaussian filter to the 3D MHD simulation with a standard deviation of 100 and 200 cells.  The simulation box is then projected into the plane of the sky in the same fashion as the previous analysis using the intensity and the velocity centroids.  Using the same procedure we obtained the Tsallis parameters and their respective coefficient of determination for the two clouds.  Even when the Tsallis parameters obtained had different values depending on the direction of the measurement, the average of their coefficient of determination was below 0.67 for the intensity, and 0.6 for the un-normalized centroid, making it overall an inadequate fit. The fact that the Tsallis parameters could not estimate the properties of the cloud might come from the fact that there is a significant different depth in each line of sight across the cloud, due to the poor computational resolution for this type of analysis.  The change in the line of sight depth also affects the projected information that translates to a change in the power spectrum measurement.  This change does not reflect changes to the turbulence but rather the LOS.

\section{Conclusions} \label{sec:conclusion}

In this paper we extended the use of the Tsallis statistic to explore the effects of anisotropy of a medium produced by the mean magnetic field.  The measurements were made in a simulated diffuse ISM with different degrees of magnetization and compressibility. The Tsallis statistic was used in the incremental PDF of the intensity and the velocity centroids. We found that the Tsallis distribution, through the Tsallis parameters, can detect the anisotropy produced by the magnetic field. We can determine the $M_s$ using the Tsallis parameters even if we do not know the relative direction of magnetic field, meaning that the Tsallis statistics does depend on magnetic field direction, but this does not compromise its use for finding the parameters of MHD turbulence that was discussed in earlier papers.  

To quantify the accuracy of the Tsallis distribution of fitting the properties of the medium we employed the coefficient of determination.  It was found that the coefficient of determination is above $\gtrsim$0.9 for most cases implying a good correlation between the Tsallis function and the data. The goodness of fit is fundamental since it determines how good the predictions are for both the direction of the mean field and the compressibility and magnetization of the media. Noise and smoothing were added to understand observational effects on the Tsallis parameters, and we find that they can be used in such conditions.   


\section*{Acknowledgements}

We thank Julie Davis and Ben Rosenwasser for insightful discussions. AL and DFGC are supported by the NSF grant DMS 1622353. Partial support for DFGC was provided by CONACyT (M\'{e}xico). AL acknowledges a distinguished visitor PVE/CAPES appointment at the Physics Graduate Program of the Federal University of Rio Grande do Norte and thanks the INCT INEspa\c{c}o and Physics Graduate Program/UFRN, at Natal, for hospitality. 



\bibliographystyle{mnras}
\bibliography{biblio-go} 

\begin{thebibliography}{}
\makeatletter
\relax
\def\mn@urlcharsother{\let\do\@makeother \do\$\do\&\do\#\do\^\do\_\do\%\do\~}
\def\mn@doi{\begingroup\mn@urlcharsother \@ifnextchar [ {\mn@doi@}
  {\mn@doi@[]}}
\def\mn@doi@[#1]#2{\def\@tempa{#1}\ifx\@tempa\@empty \href
  {http://dx.doi.org/#2} {doi:#2}\else \href {http://dx.doi.org/#2} {#1}\fi
  \endgroup}
\def\mn@eprint#1#2{\mn@eprint@#1:#2::\@nil}
\def\mn@eprint@arXiv#1{\href {http://arxiv.org/abs/#1} {{\tt arXiv:#1}}}
\def\mn@eprint@dblp#1{\href {http://dblp.uni-trier.de/rec/bibtex/#1.xml}
  {dblp:#1}}
\def\mn@eprint@#1:#2:#3:#4\@nil{\def\@tempa {#1}\def\@tempb {#2}\def\@tempc
  {#3}\ifx \@tempc \@empty \let \@tempc \@tempb \let \@tempb \@tempa \fi \ifx
  \@tempb \@empty \def\@tempb {arXiv}\fi \@ifundefined
  {mn@eprint@\@tempb}{\@tempb:\@tempc}{\expandafter \expandafter \csname
  mn@eprint@\@tempb\endcsname \expandafter{\@tempc}}}

\bibitem[\protect\citeauthoryear{{Armstrong}, {Rickett}  \&
  {Spangler}}{{Armstrong} et~al.}{1995}]{armstrong1995}
{Armstrong} J.~W.,  {Rickett} B.~J.,   {Spangler} S.~R.,  1995, \mn@doi [\apj]
  {10.1086/175515}, \href {http://adsabs.harvard.edu/abs/1995ApJ...443..209A}
  {443, 209}

\bibitem[\protect\citeauthoryear{{Audit} \& {Hennebelle}}{{Audit} \&
  {Hennebelle}}{2005}]{audit2005}
{Audit} E.,  {Hennebelle} P.,  2005, \mn@doi [\aap]
  {10.1051/0004-6361:20041474}, \href
  {http://adsabs.harvard.edu/abs/2005A%26A...433....1A} {433, 1}

\bibitem[\protect\citeauthoryear{{Blasi}}{{Blasi}}{2013}]{balsi2013}
{Blasi} P.,  2013, \mn@doi [\aapr] {10.1007/s00159-013-0070-7}, \href
  {http://adsabs.harvard.edu/abs/2013A%26ARv..21...70B} {21, 70}

\bibitem[\protect\citeauthoryear{{Burkhart}, {Falceta-Gon{\c c}alves}, {Kowal}
  \& {Lazarian}}{{Burkhart} et~al.}{2009}]{burkhart2009}
{Burkhart} B.,  {Falceta-Gon{\c c}alves} D.,  {Kowal} G.,   {Lazarian} A.,
  2009, \mn@doi [\apj] {10.1088/0004-637X/693/1/250}, \href
  {http://adsabs.harvard.edu/abs/2009ApJ...693..250B} {693, 250}

\bibitem[\protect\citeauthoryear{{Burkhart}, {Lazarian}, {Ossenkopf}  \&
  {Stutzki}}{{Burkhart} et~al.}{2013}]{burkhart2013}
{Burkhart} B.,  {Lazarian} A.,  {Ossenkopf} V.,   {Stutzki} J.,  2013, \mn@doi
  [\apj] {10.1088/0004-637X/771/2/123}, \href
  {http://adsabs.harvard.edu/abs/2013ApJ...771..123B} {771, 123}

\bibitem[\protect\citeauthoryear{{Burlaga} \& {Vi{\~n}as}}{{Burlaga} \&
  {Vi{\~n}as}}{2004}]{burlaga2004}
{Burlaga} L.~F.,  {Vi{\~n}as} A.~F.,  2004, \mn@doi [\grl]
  {10.1029/2004GL020715}, \href
  {http://adsabs.harvard.edu/abs/2004GeoRL..3116807B} {31, L16807}

\bibitem[\protect\citeauthoryear{{Chepurnov} \& {Lazarian}}{{Chepurnov} \&
  {Lazarian}}{2010}]{chepurnov2010}
{Chepurnov} A.,  {Lazarian} A.,  2010, \mn@doi [\apj]
  {10.1088/0004-637X/710/1/853}, \href
  {http://adsabs.harvard.edu/abs/2010ApJ...710..853C} {710, 853}

\bibitem[\protect\citeauthoryear{{Cho} \& {Lazarian}}{{Cho} \&
  {Lazarian}}{2002a}]{cho2002b}
{Cho} J.,  {Lazarian} A.,  2002a, \mn@doi [Physical Review Letters]
  {10.1103/PhysRevLett.88.245001}, \href
  {http://adsabs.harvard.edu/abs/2002PhRvL..88x5001C} {88, 245001}

\bibitem[\protect\citeauthoryear{{Cho} \& {Lazarian}}{{Cho} \&
  {Lazarian}}{2002b}]{cho2002}
{Cho} J.,  {Lazarian} A.,  2002b, \mn@doi [Physical Review Letters]
  {10.1103/PhysRevLett.88.245001}, \href
  {http://adsabs.harvard.edu/abs/2002PhRvL..88x5001C} {88, 245001}

\bibitem[\protect\citeauthoryear{{Cho} \& {Lazarian}}{{Cho} \&
  {Lazarian}}{2003}]{cho2003}
{Cho} J.,  {Lazarian} A.,  2003, \mn@doi [\mnras]
  {10.1046/j.1365-8711.2003.06941.x}, \href
  {http://adsabs.harvard.edu/abs/2003MNRAS.345..325C} {345, 325}

\bibitem[\protect\citeauthoryear{{Crovisier} \& {Dickey}}{{Crovisier} \&
  {Dickey}}{1983}]{crovisier1983}
{Crovisier} J.,  {Dickey} J.~M.,  1983, \aap, \href
  {http://adsabs.harvard.edu/abs/1983A%26A...122..282C} {122, 282}

\bibitem[\protect\citeauthoryear{{Draine} \& {Lazarian}}{{Draine} \&
  {Lazarian}}{1998}]{draine1998}
{Draine} B.~T.,  {Lazarian} A.,  1998, \mn@doi [\apjl] {10.1086/311167}, \href
  {http://adsabs.harvard.edu/abs/1998ApJ...494L..19D} {494, L19}

\bibitem[\protect\citeauthoryear{{Elmegreen} \& {Falgarone}}{{Elmegreen} \&
  {Falgarone}}{1996}]{elmegreen1996}
{Elmegreen} B.~G.,  {Falgarone} E.,  1996, \mn@doi [\apj] {10.1086/178009},
  \href {http://adsabs.harvard.edu/abs/1996ApJ...471..816E} {471, 816}

\bibitem[\protect\citeauthoryear{{Elmegreen} \& {Scalo}}{{Elmegreen} \&
  {Scalo}}{2004}]{elmegreen2004}
{Elmegreen} B.~G.,  {Scalo} J.,  2004, \mn@doi [\araa]
  {10.1146/annurev.astro.41.011802.094859}, \href
  {http://adsabs.harvard.edu/abs/2004ARA%26A..42..211E} {42, 211}

\bibitem[\protect\citeauthoryear{{Elmegreen}, {Kim}  \&
  {Staveley-Smith}}{{Elmegreen} et~al.}{2001}]{elmegreen2001}
{Elmegreen} B.~G.,  {Kim} S.,   {Staveley-Smith} L.,  2001, \mn@doi [\apj]
  {10.1086/319021}, \href {http://adsabs.harvard.edu/abs/2001ApJ...548..749E}
  {548, 749}

\bibitem[\protect\citeauthoryear{{Esquivel} \& {Lazarian}}{{Esquivel} \&
  {Lazarian}}{2005}]{esquivel2005}
{Esquivel} A.,  {Lazarian} A.,  2005, \mn@doi [\apj] {10.1086/432458}, \href
  {http://adsabs.harvard.edu/abs/2005ApJ...631..320E} {631, 320}

\bibitem[\protect\citeauthoryear{{Esquivel} \& {Lazarian}}{{Esquivel} \&
  {Lazarian}}{2010}]{esquivel2010}
{Esquivel} A.,  {Lazarian} A.,  2010, \mn@doi [\apj]
  {10.1088/0004-637X/710/1/125}, \href
  {http://adsabs.harvard.edu/abs/2010ApJ...710..125E} {710, 125}

\bibitem[\protect\citeauthoryear{{Falgarone}, {Lis}, {Phillips}, {Pouquet},
  {Porter}  \& {Woodward}}{{Falgarone} et~al.}{1994}]{falgarone1994}
{Falgarone} E.,  {Lis} D.~C.,  {Phillips} T.~G.,  {Pouquet} A.,  {Porter}
  D.~H.,   {Woodward} P.~R.,  1994, \mn@doi [\apj] {10.1086/174946}, \href
  {http://adsabs.harvard.edu/abs/1994ApJ...436..728F} {436, 728}

\bibitem[\protect\citeauthoryear{{Goldreich} \& {Sridhar}}{{Goldreich} \&
  {Sridhar}}{1995}]{goldreich1995}
{Goldreich} P.,  {Sridhar} S.,  1995, \mn@doi [\apj] {10.1086/175121}, \href
  {http://adsabs.harvard.edu/abs/1995ApJ...438..763G} {438, 763}

\bibitem[\protect\citeauthoryear{{Green}}{{Green}}{1993}]{green1993}
{Green} D.~A.,  1993, \mn@doi [\mnras] {10.1093/mnras/262.2.327}, \href
  {http://adsabs.harvard.edu/abs/1993MNRAS.262..327G} {262, 327}

\bibitem[\protect\citeauthoryear{{Hennebelle} \& {Ciardi}}{{Hennebelle} \&
  {Ciardi}}{2009}]{hennebelle2009a}
{Hennebelle} P.,  {Ciardi} A.,  2009, \mn@doi [\aap]
  {10.1051/0004-6361/200913008}, \href
  {http://adsabs.harvard.edu/abs/2009A%26A...506L..29H} {506, L29}

\bibitem[\protect\citeauthoryear{{Kandel}, {Lazarian}  \& {Pogosyan}}{{Kandel}
  et~al.}{2016}]{kandel2016}
{Kandel} D.,  {Lazarian} A.,   {Pogosyan} D.,  2016, \mn@doi [\mnras]
  {10.1093/mnras/stw1296}, \href
  {http://adsabs.harvard.edu/abs/2016MNRAS.461.1227K} {461, 1227}

\bibitem[\protect\citeauthoryear{{Kowal}, {Lazarian}  \& {Beresnyak}}{{Kowal}
  et~al.}{2007}]{kowal2007}
{Kowal} G.,  {Lazarian} A.,   {Beresnyak} A.,  2007, \mn@doi [\apj]
  {10.1086/511515}, \href {http://adsabs.harvard.edu/abs/2007ApJ...658..423K}
  {658, 423}

\bibitem[\protect\citeauthoryear{{Lazarian} \& {Pogosyan}}{{Lazarian} \&
  {Pogosyan}}{2000}]{lazarian2000}
{Lazarian} A.,  {Pogosyan} D.,  2000, \mn@doi [\apj] {10.1086/309040}, \href
  {http://adsabs.harvard.edu/abs/2000ApJ...537..720L} {537, 720}

\bibitem[\protect\citeauthoryear{{Lazarian} \& {Pogosyan}}{{Lazarian} \&
  {Pogosyan}}{2006}]{lazarian2006b}
{Lazarian} A.,  {Pogosyan} D.,  2006, \mn@doi [\apj] {10.1086/508012}, \href
  {http://adsabs.harvard.edu/abs/2006ApJ...652.1348L} {652, 1348}

\bibitem[\protect\citeauthoryear{{Lazarian} \& {Pogosyan}}{{Lazarian} \&
  {Pogosyan}}{2008}]{lazarian2008b}
{Lazarian} A.,  {Pogosyan} D.,  2008, \mn@doi [\apj] {10.1086/591238}, \href
  {http://adsabs.harvard.edu/abs/2008ApJ...686..350L} {686, 350}

\bibitem[\protect\citeauthoryear{{Lazarian} \& {Vishniac}}{{Lazarian} \&
  {Vishniac}}{1999}]{lazarian1999}
{Lazarian} A.,  {Vishniac} E.~T.,  1999, \mn@doi [\apj] {10.1086/307233}, \href
  {http://adsabs.harvard.edu/abs/1999ApJ...517..700L} {517, 700}

\bibitem[\protect\citeauthoryear{{Levenberg}}{{Levenberg}}{1944}]{levenberg1994}
{Levenberg} K.,  1944, Quarterly of Applied Mathematics, 2, 164

\bibitem[\protect\citeauthoryear{{Li}, {Banerjee}, {Pudritz}, {J{\o}rgensen},
  {Shang}, {Krasnopolsky}  \& {Maury}}{{Li} et~al.}{2014}]{li2014}
{Li} Z.-Y.,  {Banerjee} R.,  {Pudritz} R.~E.,  {J{\o}rgensen} J.~K.,  {Shang}
  H.,  {Krasnopolsky} R.,   {Maury} A.,  2014, \mn@doi [Protostars and Planets
  VI] {10.2458/azu_uapress_9780816531240-ch008}, \href
  {http://adsabs.harvard.edu/abs/2014prpl.conf..173L} {pp 173--194}

\bibitem[\protect\citeauthoryear{{Lis}, {Keene}, {Li}, {Phillips}  \&
  {Pety}}{{Lis} et~al.}{1998}]{lis1998}
{Lis} D.~C.,  {Keene} J.,  {Li} Y.,  {Phillips} T.~G.,   {Pety} J.,  1998,
  \mn@doi [\apj] {10.1086/306096}, \href
  {http://adsabs.harvard.edu/abs/1998ApJ...504..889L} {504, 889}

\bibitem[\protect\citeauthoryear{{Lithwick} \& {Goldreich}}{{Lithwick} \&
  {Goldreich}}{2001}]{lithwick2001}
{Lithwick} Y.,  {Goldreich} P.,  2001, \mn@doi [\apj] {10.1086/323470}, \href
  {http://adsabs.harvard.edu/abs/2001ApJ...562..279L} {562, 279}

\bibitem[\protect\citeauthoryear{{Mac Low} \& {Klessen}}{{Mac Low} \&
  {Klessen}}{2004}]{maclow2004}
{Mac Low} M.-M.,  {Klessen} R.~S.,  2004, \mn@doi [Reviews of Modern Physics]
  {10.1103/RevModPhys.76.125}, \href
  {http://adsabs.harvard.edu/abs/2004RvMP...76..125M} {76, 125}

\bibitem[\protect\citeauthoryear{{Marquardt}}{{Marquardt}}{1963}]{marquardt1963}
{Marquardt} D.~W.,  1963, Journal of the Society for Industrial and Applied
  Mathematics, 11, 431

\bibitem[\protect\citeauthoryear{{McKee} \& {Ostriker}}{{McKee} \&
  {Ostriker}}{1977}]{mckee1977}
{McKee} C.~F.,  {Ostriker} J.~P.,  1977, \mn@doi [\apj] {10.1086/155667}, \href
  {http://adsabs.harvard.edu/abs/1977ApJ...218..148M} {218, 148}

\bibitem[\protect\citeauthoryear{{McKee} \& {Ostriker}}{{McKee} \&
  {Ostriker}}{2007}]{mckee2007}
{McKee} C.~F.,  {Ostriker} E.~C.,  2007, \mn@doi [\araa]
  {10.1146/annurev.astro.45.051806.110602}, \href
  {http://adsabs.harvard.edu/abs/2007ARA%26A..45..565M} {45, 565}

\bibitem[\protect\citeauthoryear{{Miesch}, {Scalo}  \& {Bally}}{{Miesch}
  et~al.}{1999}]{miesch1999}
{Miesch} M.~S.,  {Scalo} J.,   {Bally} J.,  1999, \mn@doi [\apj]
  {10.1086/307824}, \href {http://adsabs.harvard.edu/abs/1999ApJ...524..895M}
  {524, 895}

\bibitem[\protect\citeauthoryear{{Narayan} \& {Goodman}}{{Narayan} \&
  {Goodman}}{1989}]{narayan1989}
{Narayan} R.,  {Goodman} J.,  1989, \mn@doi [\mnras] {10.1093/mnras/238.3.963},
  \href {http://adsabs.harvard.edu/abs/1989MNRAS.238..963N} {238, 963}

\bibitem[\protect\citeauthoryear{{Padoan}, {Rosolowsky}  \& {Goodman}}{{Padoan}
  et~al.}{2001}]{padoan2001}
{Padoan} P.,  {Rosolowsky} E.~W.,   {Goodman} A.~A.,  2001, \mn@doi [\apj]
  {10.1086/318378}, \href {http://adsabs.harvard.edu/abs/2001ApJ...547..862P}
  {547, 862}

\bibitem[\protect\citeauthoryear{{Padoan}, {Boldyrev}, {Langer}  \&
  {Nordlund}}{{Padoan} et~al.}{2003}]{padoan2003}
{Padoan} P.,  {Boldyrev} S.,  {Langer} W.,   {Nordlund} {\AA}.,  2003, \mn@doi
  [\apj] {10.1086/345351}, \href
  {http://adsabs.harvard.edu/abs/2003ApJ...583..308P} {583, 308}

\bibitem[\protect\citeauthoryear{{Padoan}, {Juvela}, {Kritsuk}  \&
  {Norman}}{{Padoan} et~al.}{2009}]{padoan2009}
{Padoan} P.,  {Juvela} M.,  {Kritsuk} A.,   {Norman} M.~L.,  2009, \mn@doi
  [\apjl] {10.1088/0004-637X/707/2/L153}, \href
  {http://adsabs.harvard.edu/abs/2009ApJ...707L.153P} {707, L153}

\bibitem[\protect\citeauthoryear{{Picoli Jr.}, {Mendes}, {Malacarne}  \&
  {Santos}}{{Picoli Jr.} et~al.}{2009}]{picoli2009}
{Picoli Jr.} S.,  {Mendes} R.~S.,  {Malacarne} L.~C.,   {Santos} R.~P.~B.,
  2009, \mn@doi [Brazilian Journal of Physics] {ISSN 0103-9733}, \href
  {http://dx.doi.org/10.1590/S0103-97332009000400017} {39, 439}

\bibitem[\protect\citeauthoryear{{Rosolowsky}, {Goodman}, {Wilner}  \&
  {Williams}}{{Rosolowsky} et~al.}{1999}]{rosolowsky1999}
{Rosolowsky} E.~W.,  {Goodman} A.~A.,  {Wilner} D.~J.,   {Williams} J.~P.,
  1999, \mn@doi [\apj] {10.1086/307863}, \href
  {http://adsabs.harvard.edu/abs/1999ApJ...524..887R} {524, 887}

\bibitem[\protect\citeauthoryear{{Santos-Lima}, {Lazarian}, {de Gouveia Dal
  Pino}  \& {Cho}}{{Santos-Lima} et~al.}{2010}]{santos2010}
{Santos-Lima} R.,  {Lazarian} A.,  {de Gouveia Dal Pino} E.~M.,   {Cho} J.,
  2010, \mn@doi [\apj] {10.1088/0004-637X/714/1/442}, \href
  {http://adsabs.harvard.edu/abs/2010ApJ...714..442S} {714, 442}

\bibitem[\protect\citeauthoryear{{Soler}, {Hennebelle}, {Martin},
  {Miville-Desch{\^e}nes}, {Netterfield}  \& {Fissel}}{{Soler}
  et~al.}{2013}]{soler2013}
{Soler} J.~D.,  {Hennebelle} P.,  {Martin} P.~G.,  {Miville-Desch{\^e}nes}
  M.-A.,  {Netterfield} C.~B.,   {Fissel} L.~M.,  2013, \mn@doi [\apj]
  {10.1088/0004-637X/774/2/128}, \href
  {http://adsabs.harvard.edu/abs/2013ApJ...774..128S} {774, 128}

\bibitem[\protect\citeauthoryear{{Stutzki}, {Bensch}, {Heithausen}, {Ossenkopf}
   \& {Zielinsky}}{{Stutzki} et~al.}{1998}]{stutzki1998}
{Stutzki} J.,  {Bensch} F.,  {Heithausen} A.,  {Ossenkopf} V.,   {Zielinsky}
  M.,  1998, \aap, \href {http://adsabs.harvard.edu/abs/1998A%26A...336..697S}
  {336, 697}

\bibitem[\protect\citeauthoryear{{Tofflemire}, {Burkhart}  \&
  {Lazarian}}{{Tofflemire} et~al.}{2011}]{tofflemire2011}
{Tofflemire} B.~M.,  {Burkhart} B.,   {Lazarian} A.,  2011, \mn@doi [\apj]
  {10.1088/0004-637X/736/1/60}, \href
  {http://adsabs.harvard.edu/abs/2011ApJ...736...60T} {736, 60}

\bibitem[\protect\citeauthoryear{{Tsallis}}{{Tsallis}}{1988}]{tsallis1988}
{Tsallis} C.,  1988, \mn@doi [Journal of Statistical Physics]
  {10.1007/BF01016429}, \href
  {http://adsabs.harvard.edu/abs/1988JSP....52..479T} {52, 479}

\makeatother
\end{thebibliography}


\bsp	
\label{lastpage}
\end{document}